\documentclass[aps,prl,longbibliography,twocolumn,superscriptaddress]{revtex4-2}
\usepackage{amsmath,amssymb,bm,graphicx,color,bbold,hyperref,keyval,url,latexsym,dsfont}
\usepackage{xcolor}
\usepackage[normalem]{ulem} %% \sout{text}
\usepackage{CJK}
\usepackage{float}
\usepackage{comment}

\begin{document}
\title{Ground-State-Based Model Reduction with Unitary Circuits}

\begin{CJK*}{UTF8}{}
\author{Shengtao Jiang (\CJKfamily{gbsn}蒋晟韬)}
\altaffiliation{stjiang@stanford.edu; present address: Stanford Institute for Materials and Energy Sciences, SLAC National Accelerator Laboratory and Stanford University, Menlo Park, California 94025, USA}
\affiliation{Department of Physics and Astronomy, University of California, Irvine, California 92697, USA}

\author{Steven R. White}
\affiliation{Department of Physics and Astronomy, University of California, Irvine, California 92697, USA}
\date{\today}

\begin{abstract}
We present a method to numerically obtain low-energy effective models based on a unitary transformation of the ground state. The algorithm finds a unitary circuit that transforms the ground state of the original model to a projected wavefunction with only the low-energy degrees of freedom. The effective model can then be derived using the unitary transformation encoded in the circuit.  We test our method on the one-dimensional and two-dimensional square-lattice Hubbard model at half-filling and obtain more accurate effective spin models than the standard perturbative approach.

\end{abstract}
\maketitle
\end{CJK*}

{\it Introduction.}-- Obtaining low-energy effective models (LEMs) that accurately encode the low-energy physical properties of the original models is one of the central tasks in condensed matter physics.  The LEMs typically have an exponentially smaller Hilbert space, which provides significant convenience for both analytical and numerical studies. Moreover, emergent interactions exhibited in LEMs can offer a clearer and more straightforward understanding of low-energy physical properties. For example, the pioneering derivation of antiferromagnetic superexchange~\cite{kramers1934,anderson1950} provides an intuitive understanding of antiferromagnetism in Mott insulators. 
Nowadays in condensed matter physics, perhaps the most commonly used LEMs are various spin models reduced from electronic degrees of freedom~\cite{magnetism-book}, which have achieved great success in describing low-energy excitations in materials~\cite{spinwave_lsco,Pyrochlore}. 

Typically, the derivation of LEMs utilizes a unitary transformation to decouple the designated low-energy sector from the remainder of the Hilbert space, often called a canonical transformation~\cite{ct-origin}.
An anti-Hermitian generator $S$ transforms the Hamiltonian via $H^{\rm eff}=e^{-S}He^S$, which block-diagonalizes the Hamiltonian, with the low energy sector forming the LEM. When there is an energy gap separating the two sectors, which is large compared to the couplings between them, the transformation can be carried out perturbatively. One celebrated example is the Schrieffer-Wolff transformation of the Anderson model into the Kondo model~\cite{Schrieffer-Wolff}. Another well-known case is the Hubbard model~\cite{hub-paper,hub-review1,hub-review2}:
\begin{equation}
    H=\sum_{\langle ij \rangle,\sigma}-tc^\dagger_{i\sigma} c_{j\sigma}+\sum_{i}Un_{i\uparrow}n_{i\downarrow},
\end{equation}
with $\langle ij\rangle$ denoting nearest-neighboring sites, which can be reduced to a spin-$1/2$ or $t$-$J$ model in the large $U/t$ limit by integrating out double occupancy~\cite{Takahashi_1977,macdonald}.

In contrast to various types of canonical transformations~\cite{ct-origin,ct-wegner,ct-white,ct-knetter,sasha04,ct-chan1,ct-chan2} that directly transform Hamiltonians, another frequently used but less systematic approach, facilitated by the recent development of numerical simulation, derives LEMs by matching their low-energy wavefunctions with the original models~\cite{randy,changlani,wagner1,wagner2}.  This can be useful when the canonical transformation is difficult to carry out or where it yields a large number of significant terms, provided that one only aims to match a few low-lying states instead of the whole low-energy sector.  
For example, in one approach~\cite{randy}, an effective Hamiltonian of simpler form is found that reproduces just the ground state rather than matches a spectrum. One first obtains the ground state of the original Hamiltonian, then designs an effective Hamiltonian ansatz $H(\Vec{\alpha})$ parameterized by $\Vec{\alpha}$, and uses multiple simulations to find the optimal $\Vec{\alpha}$ that maximizes the overlap of the new and original ground state.
However, the optimization cost can be high if the ansatz contains many parameters and a large number of simulations are required. 
In some cases, the original Hamiltonian simulation may identify degrees of freedom as being unoccupied, allowing a very simple truncation.  For example, in a recent work on the three-band Hubbard models, DMRG simulations revealed that many one-particle orbitals were almost completely unoccupied, allowing them to be truncated without the need for two-particle unitary transformations~\cite{downfolding-our}.  The LEM was then formed by a simple single particle rotation to the Wannier functions spanning the occupied orbitals. However, in the general case, when the higher bands have finite occupancy or when the high-energy degrees of freedom exhibit many-body effects such as significant double occupancy, a many-body unitary transformation is required to derive LEMs.

In this letter, we propose a systematic and efficient method to obtain LEMs based on a unitary transformation of the ground state. The algorithm is based on tensor networks, which find a proper unitary circuit $G$ made of local gates that transforms the ground state of the original model to a projected wavefunction with only low-energy degrees of freedom. 
The effective model can then be generated via $H^{\rm eff}=GHG^\dagger$, whose terms are local and few-body by the design of the unitary circuit, with their coefficients insensitive to system sizes. It is highly efficient since no more simulations other than obtaining the ground state of the original model are required. It is also non-perturbative and derives more accurate LEMs compared to the standard perturbative approach for intermediate coupling. The validity of the generated LEMs can be verified by simulation of their low-lying states.

%-------------------------------------------------------------
\begin{figure}[t]
	\includegraphics[width=0.7\columnwidth]{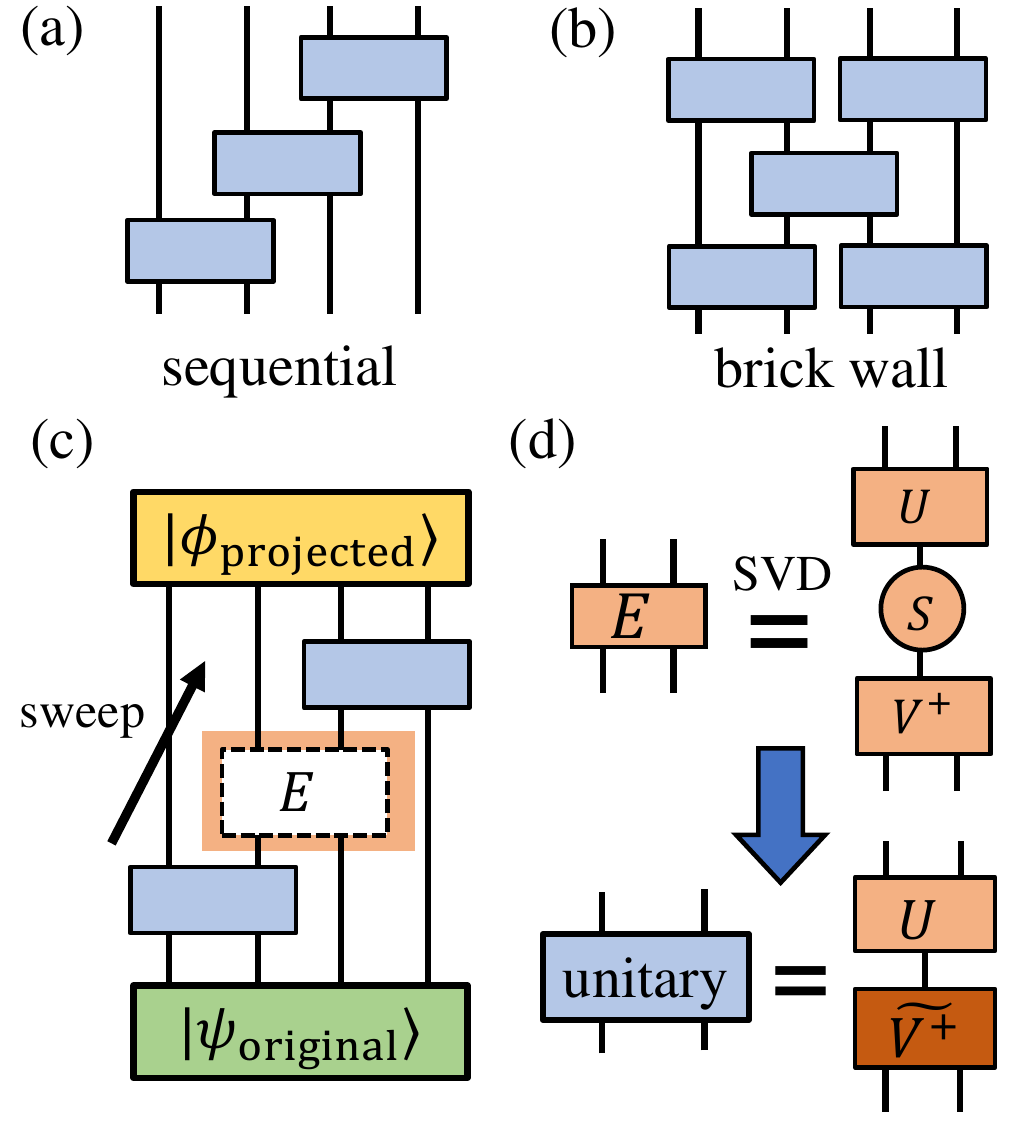}
	\caption{Two structures of the unitary circuits considered in this work: (a) a sequential circuit, and (b) a brick wall circuit. 
    (c) To find the local unitaries that transform $|\psi_{\rm original}\rangle$ to $|\phi_{\rm projected}\rangle$, we optimize one unitary at a time with the others fixed and sweep through the whole circuit back and forth. The optimal individual unitary is constructed based on the environmental tensor $E$, which is obtained by leaving out the corresponding unitary in the contraction of the circuit with wavefunctions. (d) To optimize each unitary, one performs an SVD of the tensor $E$, adjusts the tensor $V^\dagger$ into $\Tilde{V^\dagger}$, and recombines with the tensor $U$ to form the optimal unitary gate.}
	\label{fig:algorithm}
	\vskip -0.4cm
\end{figure}
%-------------------------------------------------------------

{\it Algorithm.}--- The algorithm begins by using DMRG~\cite{dmrg1,dmrg2,itensor} to simulate the original Hamiltonian $H^{0}$ to obtain its ground state $|\psi_{0}\rangle$. We then project out certain high-energy degrees of freedom and get $|\phi_{P}\rangle=\hat{P}|\psi_{0}\rangle$ (with normalization). In the Hubbard model, which is our main focus, we take $\hat{P}$ as the Gutzwiller projection~\cite{Gutzwiller}, which projects out double occupancy. In a multi-band model, one could choose a projection operator that removes higher bands of single-particle states. 
The goal of the projection is to identify the target wavefunction with only the effective low-energy degrees of freedom. Consider a two-site half-filled Hubbard model as an example. Its ground state is $|\psi_{0}\rangle =c_1(|\uparrow, \downarrow \rangle-|\downarrow, \uparrow \rangle) + c_2(|0, \uparrow\downarrow \rangle +  |\uparrow\downarrow, 0 \rangle)$.  We would like the ground state of the LEM $|\phi_P\rangle$ to omit the doubly-occupied states and retain the same singlet form as the original $|\psi_0\rangle$, i.e., $|\phi_P\rangle \sim c_1(| \tilde{\uparrow}, \ \tilde{\downarrow} \rangle-|\tilde{\downarrow}, \tilde{\uparrow} \rangle)$. Therefore, $|\phi_P\rangle$ is just the Gutzwiller projection of $|\psi_0\rangle$ where the spins are the effective degrees of freedom. 
We note that the primary use of the projected state is to find an appropriate unitary transformation $G$ satisfying $G|\psi_{0}\rangle=|\phi_{P}\rangle$. In particular, we are not interested in the energy of the projected state using the original Hamiltonian.  The LEM is generated via $H^{\rm eff}=GHG^\dagger$, which guarantees its ground state to be $|\phi_{P}\rangle$.

Since $|\psi_{0}\rangle$ and $|\phi_{P}\rangle$ are only two states in the exponentially large Hilbert space, there are many unitary transformations $\hat{G}$ that satisfy $\hat{G}|\psi_{0}\rangle=|\phi_{P}\rangle$ while transforming other states differently. To obtain a $\hat{G}$ that can produce a useful $H^{\rm eff}$ with local and few-body terms, it needs to be constructed with local and few-body unitary transformations as well, $\hat{G}=\prod_i\hat{g_i}$.   Therefore, a natural ansatz for $\hat{G}$ is a unitary circuit (also known as quantum circuit).  In this paper, we have considered two different structures of unitary circuits, the sequential circuit and the brick wall circuit, shown in Fig.~\ref{fig:algorithm}(a) and (b), respectively.   The sequential circuit turns out to be more useful as it can achieve a higher fidelity between the transformed state and the target state (with a similar number of gates), potentially due to its capacity to generate long-ranged correlations~\cite{prxq-lin}.

After the structure of the circuit is chosen, we optimize the unitaries to maximize $\langle\phi_{P}|\hat{G}|\psi_{0}\rangle$. Here, we optimize one unitary at a time with the other unitaries fixed, and sweep through the whole circuit back and forth, as shown in Fig.~\ref{fig:algorithm}(c). This ``DMRG-like" local optimization has also been studied in various works with good performance~\cite{prx-haghshenas,prxq-lin,evenbly_vidal}. 

%-------------------------------------------------------------
\begin{figure*}[t]
	\centering
\includegraphics[width=1.5\columnwidth]{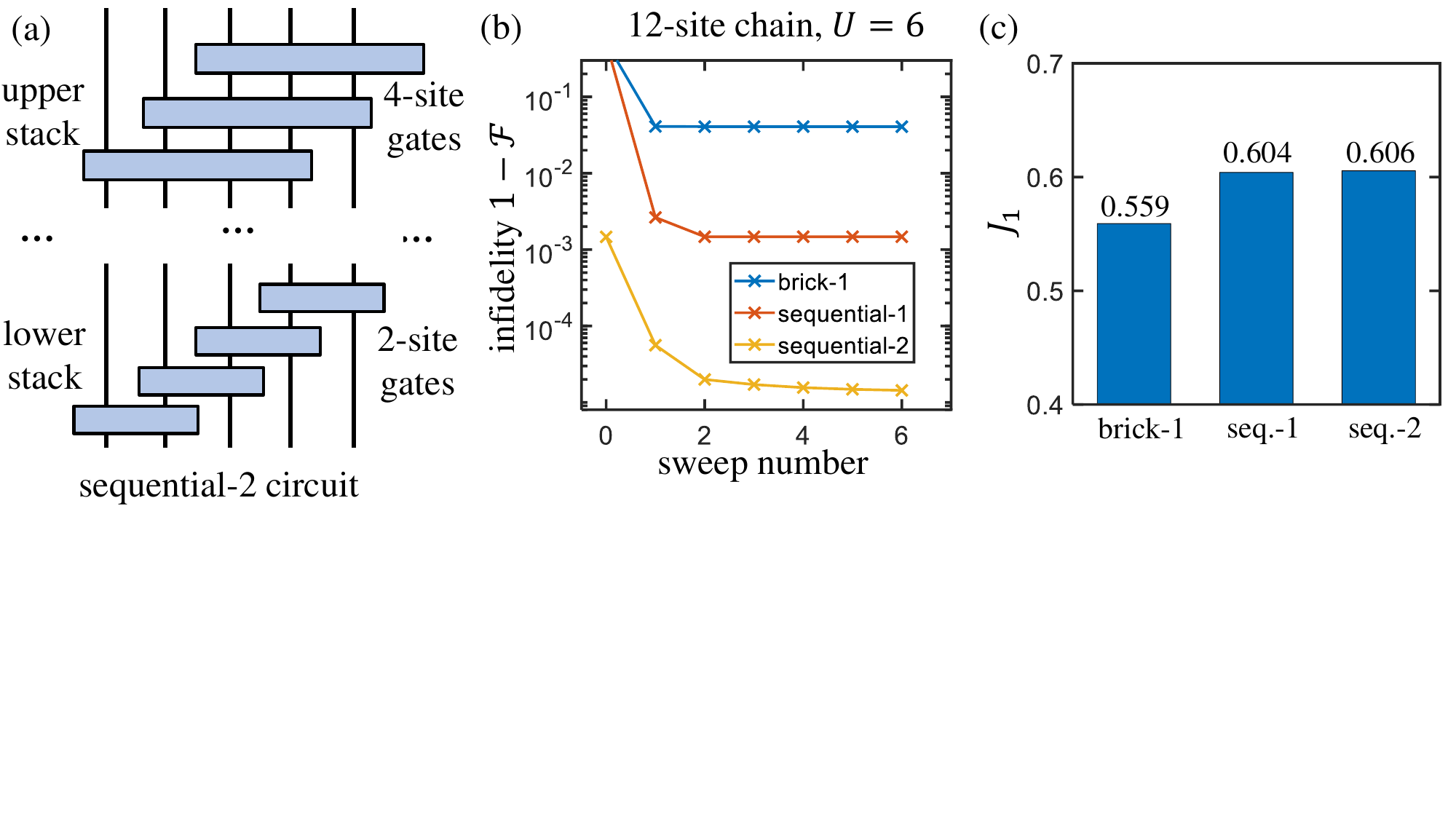}
	\vskip -3.0cm
	\caption{(a) For a 1D chain, the sequential-2 circuit is made of sequential 2-site gates in the lower stack and sequential 4-site gates in the upper stack. (b) For a 12-site Hubbard chain with $U$=6, the infidelity 1-$\mathcal{F}$ reached by different circuits. (c) The nearest-neighbor spin exchange $J_1$ extracted from different circuits.}
	\label{fig:chain}
	\vspace{-0.0cm}
\end{figure*}
%-------------------------------------------------------------

At each step, the optimal unitary is obtained using the Evenbly-Vidal algorithm~\cite{evenbly_vidal}, but with additional steps to avoid encoding unnecessary transformations in the unitaries. As shown in Fig.~\ref{fig:algorithm}(c) and (d), we start by contracting the whole circuit with the two states except the desired gate, to obtain the environmental tensor $E$.  We perform a singular value decomposition (SVD) on $E$: 
\begin{equation}
	\label{eq:svd}
	E=\sum_i S_i |U_i\rangle \langle V_i|,   
\end{equation}
The unitary $\hat{g}$ maximizing the overlap is constructed by removing the singular values $S_i$: 
\begin{equation}
	\hat{g}= \sum_i|U_i\rangle \langle V_i|.  
\end{equation}
Note that for $\hat{g}$ to be a unitary, both $\{|U_i\rangle\}$ and $\{|V_i\rangle\}$ need to be a complete orthonormal basis set. 
However, not only are there vanishing individual singular values, but there are entire blocks that vanish. For example, if $|\phi_{P}\rangle$ has double occupancy projected out, then the relevant blocks in $E$ will be zero. Any vectors $\{|\hat{U}_j\rangle\}$ and $\{|\hat{V}_j\rangle\}$ with zero singular values are not uniquely determined by the SVD in Eq.~\ref{eq:svd}.
While these different unitaries can achieve the same overlap (which is solely determined by the unique part in SVD with finite $S_i$), the derived LEMs will be different.
To maximally preserve the locality and few-body nature of the original model, we need to minimize the rotation encoded in this flexible part of the unitary.  To achieve this, the $\{|\hat{U}_j\rangle\}$ and $\{|\hat{V}_j\rangle\}$ need to be rotated to match as much as possible.  Inspired by the procedure in generating Wannier functions~\cite{downfolding-our}, for each $|U_j\rangle$ in the zero-singular value space, we make the linear combination of the $\{|V_j\rangle\}$ closest to $|U_j\rangle$:
\begin{equation} 
	W_j=\sum_{j'}\langle \hat{V}_{j'}|\hat{U}_{j}\rangle|\hat{V}_{j'}\rangle,
\end{equation}
then symmetrically orthonormalize~\cite{lowdin1950} $\{W_j\}$ to obtain $\{|\Tilde{V_j}\rangle\}$. This subset updates the complete basis $\{|V_i\rangle\}$ to $\{|\Tilde{V}_i\rangle\}$. The resulting proper unitary is:
\begin{equation}
	\hat{g}= \sum_i|U_i\rangle \langle \Tilde{V}_i|.
\end{equation}
In practice, we set a cutoff of $10^{-5}$ for $S_i$ below which we implement the above scheme for the corresponding $|\hat{U}_i\rangle$ and $|\hat{V}_i\rangle$. This has a minimal effect on the overlap but can avoid some unnecessary transformations being encoded in the unitaries.

We find that a modest number of sweeps (5-10, see Fig~\ref{fig:chain}(b)) are sufficient to make the transformed state converge to the target state. Then the LEM can be generated with $H^{\rm eff}_{\rm MPO}=GHG^\dagger$.  As a tensor network, this is a matrix product operator with additional unitary gates on both sides, which can be simplified to a single matrix product operator. 

We seek to convert the MPO $H^{\rm eff}_{\rm MPO}$ to an analytic form as a sum of local terms, which we call $H^{\rm eff}(\vec{h})$ with $\vec{h}$ denoting the magnitude of the terms. There are subtleties associated with this process. 
For example, let us consider extracting a spin exchange between sites 1 and 2 in a fermionic model. One way is to contract the MPO tensors from sites 3 to N with an identity operator, which projects out the terms involving other sites.  
However, the above procedure does not work if the MPO has a term $S^z_1 S^z_2 n_3$, which is essentially the same as $S^z_1S^z_2$ in the low-energy spin space ($\langle n_3 \rangle=1$) of the half-filled Hubbard model we will be dealing with later. 
Therefore, it makes sense to extract the Hamiltonian terms under a basis that resides in the low-energy sector, i.e., to set a spin background, which can be done in an unbiased way using perfect sampling~\cite{perfectsampling} of the projected ground state $|\phi_{P}\rangle$ to obtain a product state of spins.

With the appropriate background, we can proceed to extract the terms. The effects of different terms can be mixed together. For example, both $J_1$ and $J_c$ (defined in Eq.~\ref{eq:heffj1j2jc}) can cause a pair of spin flips. Therefore, we evaluate $H^{\rm eff}_{\rm MPO}$ on a cluster instead of a single bond, such that multiple terms are supported and considered simultaneously. 
The detailed procedure is: 
(1) obtain a product state $|\alpha\rangle$ by sampling the projected ground state $|\phi_{P}\rangle$, 
(2) choose a cluster (e.g., a 2$\times$2 plaquette) and construct a basis $\{\alpha_i\}$ by listing all the spin configurations of cluster with the background spins outside unchanged as $|\alpha\rangle$,
(3) compute all the matrix elements of $\langle\alpha_{i'}|H^{\rm eff}_{\rm MPO}|\alpha_{i}\rangle$. 
(4) find the analytic $H^{\rm eff}(\vec{h})$ that reproduces those matrix elements: $\langle\alpha_i|H^{\rm eff}(\vec{h})|\alpha_{i'}\rangle=\langle\alpha_i|H^{\rm eff}_{\rm MPO}|\alpha_{i'}\rangle$~\footnote{To search for the analytic $H^{\rm eff}(\Vec{h})$, one can start with an initial guess $H^{\rm eff}(\Vec{h_0})$, potentially from perturbation theory, and adjust the parameters as well as add or subtract terms to match $H^{\rm eff}_{\rm MPO}$.}. 
This procedure is repeated with multiple samplings, as well as varying the size and position of the selected cluster. If all of them yield the same $H^{\rm eff}(\Vec{h})$, then the conversion from $H^{\rm eff}_{\rm MPO}$ to an analytic form $H^{\rm eff}(\vec{h})$ can be confirmed.

\begin{figure}[t]
	\centering
    \includegraphics[width=1.0\columnwidth]{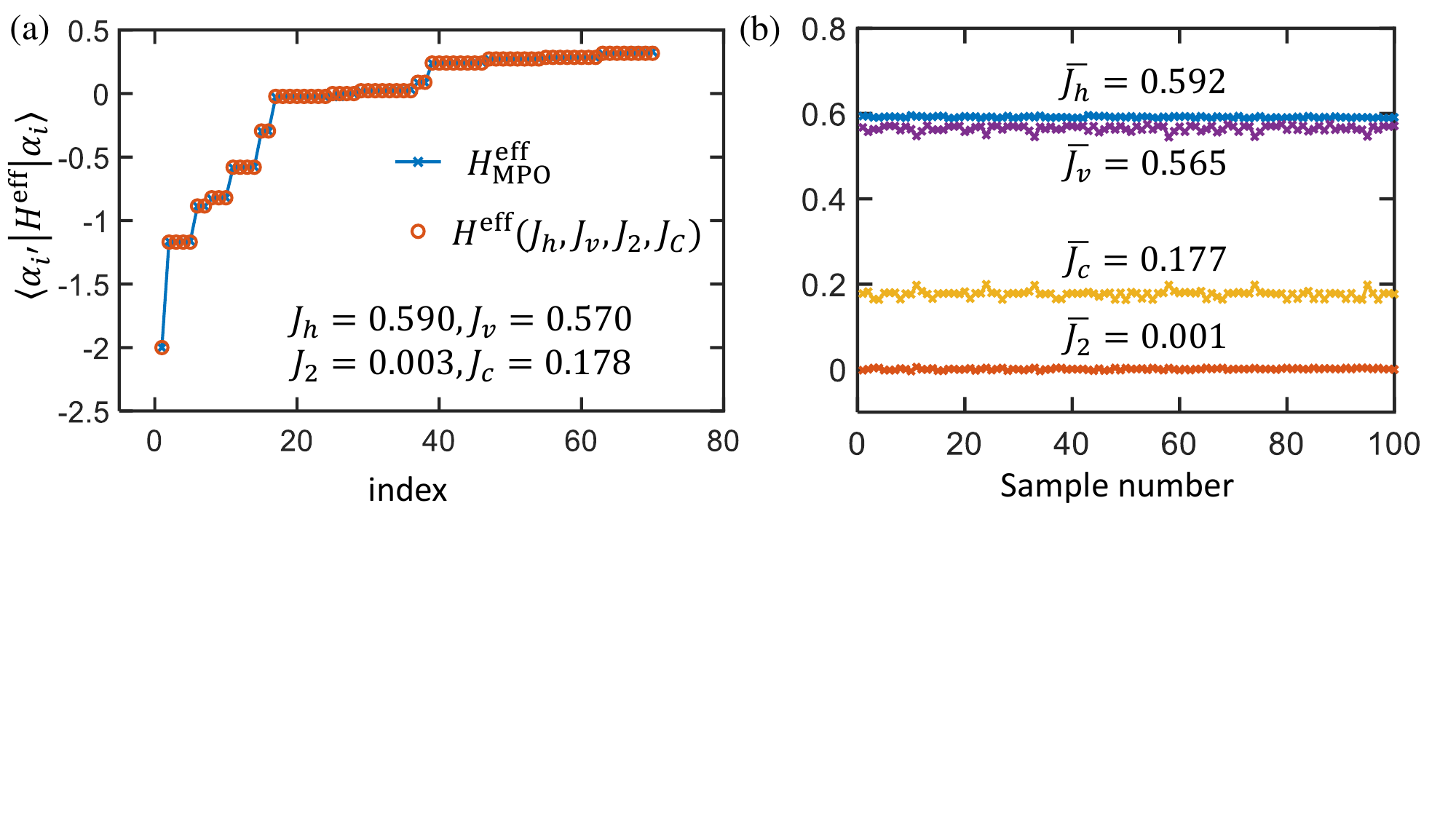}
	\vskip -2.0cm
	\caption{Converting the numerical $H^{\rm eff}_{\rm MPO}$ to an analytic $H^{\rm eff}(\vec{h})$ for the 2$\times$12 Hubbard ladder at $U$=6 in Fig.~\ref{fig:ladder} ($t$=1 as the energy unit). (a) The 70 nonzero matrix elements of $\langle\alpha_{i'}|H^{\rm eff}_{\rm MPO}|\alpha_{i}\rangle$ and their corresponding $\langle \alpha_{i'}|H^{\rm eff}(J_h,J_v,J_2,J_c)|\alpha_i\rangle$, sorted in ascending order. The close match indicate $H^{\rm eff}_{\rm MPO}$ can be well-described by these four spin exchanges with almost nothing left out. All the diagonal matrix elements $\langle\alpha_{i}|H^{\rm eff}_{\rm MPO}|\alpha_{i}\rangle$ and $\langle \alpha_{i}|H^{\rm eff}(J_h,J_v,J_2,J_c)|\alpha_i\rangle$ are shifted by a rough total energy estimate $\sim-N(J_h+J_v)/2$ to fit in the plot. 
    $\{|\alpha_i\rangle\}$ is a basis set constructed from one sampled product state $|\alpha\rangle$ (see text).
    (b) The extracted $(J_h,J_v,J_2,J_c)$ for each sample and the average values $(\bar J_h,\bar J_v,\bar J_2,\bar J_c)$ over all samples.}
	\label{fig:extract}
\end{figure}

An example of converting $H^{\rm eff}_{\rm MPO}$ to $H^{\rm eff}(\vec{h})$ is presented in Fig.~\ref{fig:extract}. Here, the sampled product state $|\alpha\rangle$ is the N\'eel state, based on which the basis $\{\alpha_i\}$ is constructed by changing the spin configuration of the central 2$\times$2 plaquette.
Fig.~\ref{fig:extract}(a) shows that all the matrix elements of $H^{\rm eff}_{\rm MPO}$ under this basis match well with a Hamiltonian containing four types of spin exchanges $H^{\rm eff}(J_h, J_v, J_2, J_c)$ (defined in Eq.~\ref{eq:heffj1j2jc}, $J_h$ and $J_v$ are $J_1$ interaction on horizontal and vertical bonds, which are approximately the same), with an average difference of only 0.0059.
Also, the extracted $\vec{h}=(J_h,J_v,J_2,J_c)$ only varies slightly between different samples as shown in Fig.~\ref{fig:extract}(b).
We further checked different cluster shapes and found no significant longer-ranged term. The extracted exchanges also do not depend on the position of the selected cluster. These evidences confirm that the extracted analytic $H^{\rm eff}(J_h,J_v,J_2,J_c)$ is a good approximation of the original $H^{\rm eff}_{\rm MPO}$ from the unitary circuit.

%-------------------------------------------------------------
%------------------------------------------------------

%-------------------------------------------------------------
\begin{figure*}[ht]
	\centering	\includegraphics[width=1.3\columnwidth]{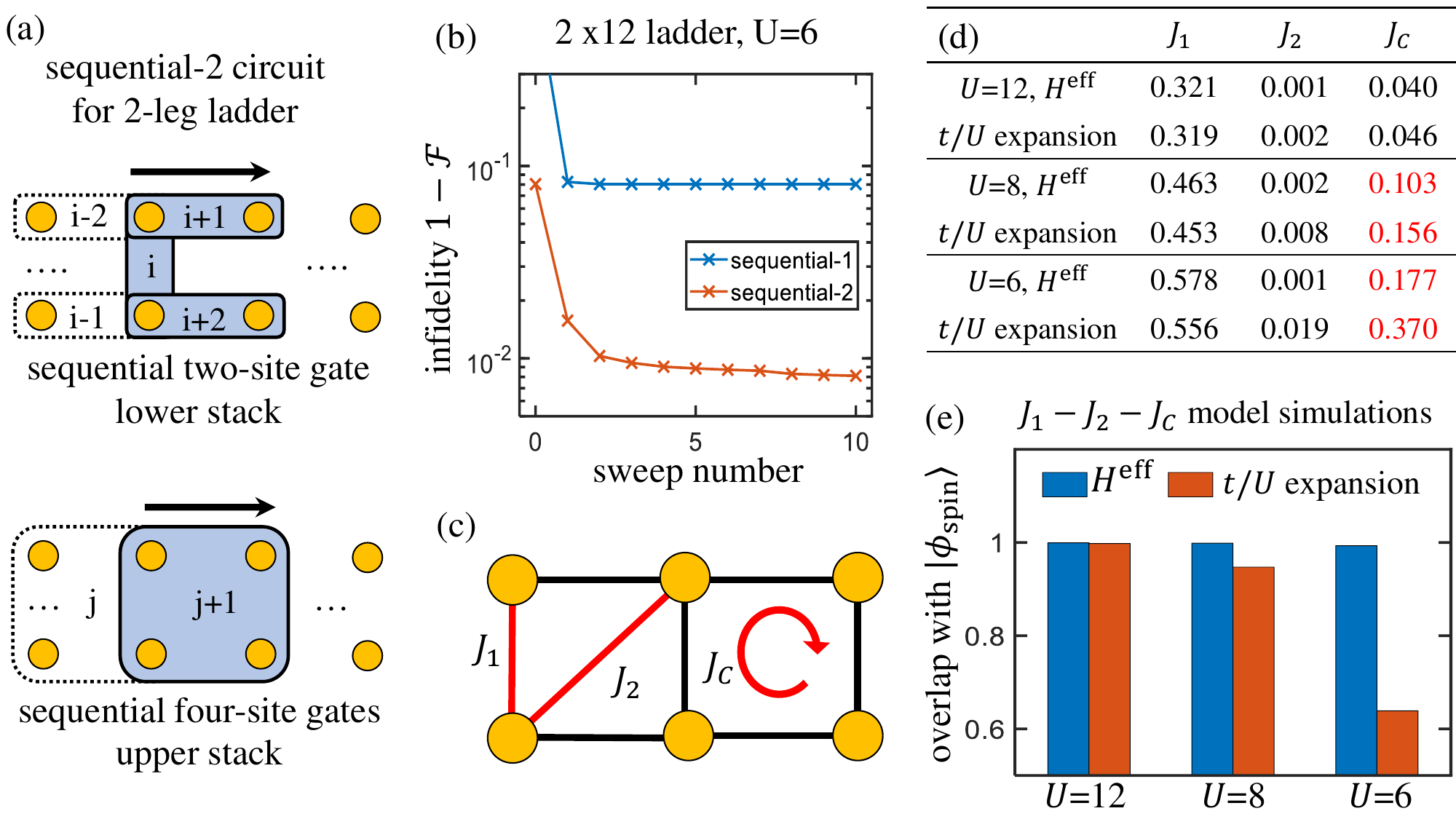}
	\vskip -0.3cm
	\caption{(a) The structure of the sequential-2 circuit for a 2-leg ladder: the lower two-site gates stack has three gates per rung, while the upper four-site gates stack has one gate per rung. The numbers on the gates indicate their orders in the circuit. 
    For a 2$\times$12 Hubbard ladder at $U$=6, (b) the infidelity 1-$\mathcal{F}$ reached by different circuits.
    (c) Illustration of spin exchanges in the effective $J_1$-$J_2$-$J_c$ model. 
    (d) The magnitude of the spin exchanges extracted from $H^{\rm eff}$ compared with those from the fourth-order $t/U$ expansion.
    (e) Squared overlap between the ground states of the $J_1$-$J_2$-$J_c$ model obtained from DMRG simulations and the original projected wavefunction $|\phi_{\rm spin}\rangle$, using parameters extracted from $H_{\rm eff}$ and from the $t/U$ expansion.} 
	\label{fig:ladder}
	\vspace{-0.0cm}
\end{figure*}
%-----------------------------------------------------------

{\it Results.}---We first test our algorithm on a half-filled 12-site Hubbard chain at an intermediate coupling $U=6$ ($t=1$ is set as the energy unit throughout the paper). The results are presented in Fig.~\ref{fig:chain}. The simple sequential-1 circuit made of two-site gates only, already reduces the infidelity $1-\mathcal{F}:=1-|\langle\phi_{P}|G_{\rm seq1}|\psi_{0}\rangle|^2$ to $\mathcal{O}(10^{-3})$ within 2 sweeps. 
To improve upon it, we have tried to expand the circuit with 2-site gates, either by replacing each gate with three sequential gates or by adding another stack. However, these structures generally do not offer a substantial improvement, suggesting the limitation of local 2-site gates. Therefore, a further improved circuit, namely the sequential-2 circuit, adds a stack of 4-site gates on top of the sequential-1 circuit to deal with longer-ranged and four-body interactions~\footnote{Here we skip using 3-site gates because (1) the improvement is not as big as using 4-site gates possibly due to lack of three-body interactions; (2) 4-site gates are a more natural choice for dealing with two-dimensional square lattice in the later part.}.  When optimizing the 4-site gates, the lower stack of 2-site gates is fixed to avoid absorbing the 2-site transformation into the 4-site ones.
On the other hand, we have also tried the brick-1 circuit, which has one layer of 2-site gates on odd bonds followed by another layer on even bonds. Its infidelity is one magnitude bigger, possibly because of its inability to generate long-range correlation, similar to the observation in Ref.~\cite{prxq-lin}. Furthermore, the $H^{\rm eff}$ it yields differs between even and odd bonds. For these reasons, we consider the sequential circuit to be a better choice. 

For the resulting $H^{\rm eff}(\vec{h})$ for the 1-D chain, the only significant interaction is the nearest-neighbor (NN) Heisenberg exchanges $J_1$, while the next-NN and four-body exchange coefficients are $\lesssim 0.01$. This is verified by a simulation of the Heisenberg chain, whose ground state is almost identical to that of the original Hubbard ground state projected into spin space, with an infidelity of only $1.6\times10^{-4}$.
Alternatively, $H^{\rm eff}_{\rm MPO}$ as an MPO has a bond dimension of 16, since it also contains interaction involving doubly occupied or empty states. After projecting out those states and applying a cutoff of 1E-4, its bond dimension reduces to 5, the same as a Heisenberg chain.
In Fig.~\ref{fig:chain}(c), we show the extracted $J_1$ from different circuits, which converges with the increase of the fidelity.  Matching the spin gap between the Hubbard chain and the Heisenberg model also yields $J_1\approx$ 0.60. This is expected since all the energy levels are preserved in a unitary transformation of the Hamiltonian. 

Next, we test our algorithm on a two-leg Hubbard ladder. The sequential circuit is shown in Fig.~\ref{fig:ladder}(a). 
Its structure resembles a staircase: starting from the left end of the system, a gate is acted on the rung, followed by two gates acting on the legs, and this pattern continues to the right.
The infidelity is further reduced by an order of magnitude by adding another stack of four-site gates on top of the two-site gates, as shown in Fig.~\ref{fig:ladder}(b). 

For the two-leg ladder, there are four types of spin exchange terms ($J_1$, $J_2$, $J_3$, $J_c$) according to the fourth order $t/U$ perturbative expansion: $J_1=4t^2/U-24t^4/U^3$, $J_2=J_3=4t^4/U^3$, $J_c=80t^4/U^3$~\cite{macdonald,Takahashi_1977}.  We first extract those acting within a two-by-two plaquette as illustrated in Fig~\ref{fig:ladder}(c).  We also checked $J_3$ and other long-ranged terms, which are found to be less than 0.02. The form of the approximate effective Hamiltonian $H^{\rm eff}(J_1,J_2,J_c)$ is:
\begin{equation}
    \label{eq:heffj1j2jc}
	\begin{aligned}
		&H^{\rm eff}(J_1,J_2,J_c)=
		J_1\sum_{\langle ij \rangle_1}{{\bf S}_i}\cdot{{\bf S}_j} +
		J_2\sum_{\langle ij \rangle_2} {{\bf S}_i}\cdot{{\bf S}_j} \\
		&+J_c\sum_{\langle ijkl \rangle}({{\bf S}_i} \cdot{{\bf S}_j})({{\bf S}_k} \cdot{{\bf S}_l})
		+({{\bf S}_i} \cdot{{\bf S}_l})({{\bf S}_j} \cdot{{\bf S}_k}) \\
		&\hspace{1.1cm} -({{\bf S}_i} \cdot{{\bf S}_k})({{\bf S}_j} \cdot{{\bf S}_l}),
	\end{aligned}
\end{equation}
where $\langle ij \rangle_{1(2)}$ denotes the first(second) nearest neighbor pair of sites and $\langle ijkl \rangle$ denotes a 2$\times$2 plaquette.
The process of extracting spin exchanges is shown in Fig.~\ref{fig:extract} as mentioned previously. 
Averaging the NN horizontal exchange $J_h$ and vertical exchange $J_v$ to $J_1$,  
the resulting $(J_1, J_2, J_c)$ at three different coupling $U$ are listed in Fig.~\ref{fig:ladder}(e),  and compared with those obtained from the fourth-order $t/U$ expansion. As expected, the parameters match well at strong coupling $U=12$. However, at intermediate couplings $U=8$ and $U=6$, while the $J_1$ remains similar for both approaches, the four-spin cyclic exchange term $J_c$ is much smaller in $H^{\rm eff}$. 
To test the validity of the obtained $H^{\rm eff}(J_1, J_2, J_c)$, we simulate it using DMRG and compute the overlap of its ground state with the original projected Hubbard wavefunction. As shown in Fig.~\ref{fig:ladder}(e), $H^{\rm eff}(J_1,J_2,J_c)$ produces essentially the same ground state (infidelity $<$0.01) for all three cases, while the $t/U$ expansion becomes much less accurate as $U$ decreases.

We further tested our method in two more challenging systems. The first one is a $4\times8$ cylinder at half-filling with $U=8$. In this case, with a fidelity of 0.977 between the transformed wavefunction and the targeted one,  the $J_c$ extracted from $H^{\rm eff}$ is only around half of that from the $U/t$ expansion, similar to the two-leg ladder case.  We note that the spin exchanges obtained ($J_1$=0.437, $J_2$=-0.004, $J_c$=0.083) on four-leg cylinder is similar to that in the two-leg ladder case, indicating the validity and generality of the effective model derived on relatively small cluster.
Secondly, we apply our algorithm to a doped two-leg ladder with $U$=7.5 and doping=0.125 and obtain an effective $t$-$J$ model.  In addition to the various spin exchanges, we extract a three-site correlated hopping term~\cite{anderson1959} with $t_{\rm 3-site}/t$=0.11, comparable to the $t_{\rm 3-site}/t$=0.13 from the second order $t/U$ expansion. However, in the doped case $H^{\rm eff}$ is not as well approximated by these standard terms as the undoped case.  Although these cases are more challenging, the results support the algorithm's applicability to both two-dimensional and doped models.

{\it Discussions and Summary.}---
In the derivation of LEMs, ideally, one wishes to find an effective Hamiltonian that fully captures a certain low-energy sector of the original model in terms of both energy levels and the corresponding states. The unitary transformation preserves the energy levels. Although in our algorithm, there is no guarantee that the excited states will also match, in our test on the two-leg half-filled Hubbard ladder at $U=6$ (Fig.~\ref{fig:ladder}), the first excited state also matches with nearly the same overlap as the ground state.
One future improvement is to require the unitary circuit to transform not only the ground states but also several low-lying excited states, potentially utilizing state-averaged MPS.

Regarding the structure of the unitary circuit,  we have explored removing the two-site gates and using only four-site gates for the circuit in Fig.~\ref{fig:ladder}. While a similar fidelity between the transformed wavefunction and the targeted one can be reached, the $H^{\rm eff}$ extracted from the circuit can not be as well-approximated by a simple $H^{\rm eff}(J_1, J_2, J_c)$, with the average difference in the Hamiltonian elements around six times larger than that from the sequential-2 circuit in Fig.~\ref{fig:ladder}(a). This confirms that the circuit should be constructed primarily using local and few-body gates, with larger gates employed only to transform the remaining parts that are beyond the reach of the smaller ones. This ensures that the resulting $H^{\rm eff}$ remains local and few-body.

We note that an alternative approach that utilizes ground states to derive LEMs is the parental Hamiltonian methods~\cite{parent-clark,parent-grover,parent-yang,parent-qi}. In this case, one directly optimizes $H^{\rm eff}$ such that the given target state becomes its eigenstate.  In these approaches, one needs to take an extra step to determine the scale of $H^{\rm eff}$, which can be done by matching the excitation gap.

In summary, we propose a tensor-network-based method to derive LEMs based on a unitary transformation of the ground state. The algorithm finds a proper unitary circuit $G$ made of local gates that integrates out the high-energy degrees of freedom in the ground state, which is then used to derive the effective model via $H^{\rm eff}=GHG^+$. The resulting effective model is general and largely insensitive to system size.. The algorithm is highly efficient and only requires a single simulation to obtain the ground state of the original model. Its effectiveness is tested on the square-lattice Hubbard model at half-filling and shown to generate more accurate LEMs compared to the perturbative approach at intermediate coupling.

{\it Data Availability.}---
The data used to generate the figures are deposited in Ref.~\footnote{\href{https://doi.org/10.6084/m9.figshare.28830341}{https://doi.org/10.6084/m9.figshare.28830341}}.

%----------------------------------------
\begin{acknowledgments}
\emph{Acknowledgments.}---
SJ and SRW are supported by the NSF under DMR-2412638. 
\end{acknowledgments}
%----------------------------------------

\bibliography{ref}
\end{document}